# Observing a quantum Maxwell demon at work

N. Cottet[1*], S. Jezouin[1*], L. Bretheau[1]†, P. Campagne-Ibarcq[1]‡, Q. Ficheux[1], J. Anders[2], A. Auffèves[3], R. Azouit[4,5], P. Rouchon[4,5] and B. Huard[1,6]

[1]Laboratoire Pierre Aigrain, Ecole normale supérieure, PSL Research University, CNRS, Université Pierre et Marie Curie, Sorbonne Universités, Université Paris Diderot, Sorbonne Paris-Cité, 24 rue Lhomond, 75231 Paris Cedex 05, France.
[2]University of Exeter, Stocker Road, Exeter EX4 4QL, United Kingdom.
[3]Institut Néel, UPR2940 CNRS and Université Grenoble Alpes, avenue des Martyrs, 38042 Grenoble, France
[4]Centre Automatique et Systèmes, Mines ParisTech, PSL Research University, 60 Boulevard Saint-Michel, 75272 Paris Cedex 6, France.
[5]Quantic Team, INRIA Paris-Rocquencourt, Domaine de Voluceau, B.P. 105, 78153 Le Chesnay Cedex, France.
[6]Laboratoire de Physique, Ecole Normale Supérieure de Lyon, 46 allée d'Italie, 69364 Lyon Cedex 7, France.

*these authors contributed equally to the work presented
†current address: Department of Physics, Massachusetts Institute of Technology, Cambridge, Massachusetts 02139, USA.
‡current address: Department of Applied Physics and Physics, Yale University, New Haven, Connecticut 06520, USA.

**In 1867, pondering the newly developed thermodynamic laws, Maxwell came to the disturbing conclusion that a "demon" can extract work cyclically from a thermodynamic system beyond the limits set by the second law when acting upon the information it obtains about the system[1]. This paradox was resolved a century later when Landauer and Bennett realized the crucial thermodynamic role of the information stored in the demon's memory[2]. With the advent of quantum information, the behavior of Maxwell demons in the quantum regime arose strong interest[3–7]. While recent experiments have realized classical versions of elementary Maxwell demons in various physical systems[8–13], experimental realizations in the quantum regime are in their infancy[13] and a full characterization is still missing. Here we reveal the inner mechanics of a quantum Maxwell demon that is able to extract work from a quantum system using superconducting circuits. Importantly, we are able to directly probe the extracted work by measuring the output power emitted by the system at the single photon level, without inferring it from system trajectories[8–11,14]. We are thus able to demonstrate how the information stored in the demon's memory affects the extracted work. To make the characterization complete, we also measure the entropy and energy of the system and the demon. Superconducting circuits thus reveal themselves as a suitable experimental testbed for the blooming field of quantum thermodynamics of information[15–19].**

The experimental exploration of quantum thermodynamics requires quantum systems whose states and energy flows can be fully measured. Recently, superconducting circuits have been shown to provide such a level of control. The basic components of a Maxwell's demon experiment are a system, a memory that can store information on the system and a battery in which the demon can release the work it extracts from the system using the information stored in the memory. In this experiment the demon is a microwave cavity that encodes quantum information about a superconducting qubit and converts information into work by powering up a propagating microwave pulse by stimulated emission.

The system S is thus a transmon superconducting qubit [20] with energy difference $hf_S = h \times 7.09$ GHz between its ground $|g\rangle$ and excited $|e\rangle$ states. It is embedded in a microwave cavity, the demon D, that resonates at $f_D = 7.91$ GHz. The dispersive Hamiltonian reads $H = hf_S|e\rangle\langle e|_S + hf_D d^\dagger d - h\chi d^\dagger d|e\rangle\langle e|_S$, where $d$ is the annihilation operator of a photon in the cavity. The last term induces a frequency shift of the cavity by $-\chi = -33$ MHz when the qubit is excited. Reciprocally the qubit frequency is shifted by $-N\chi$ when the cavity hosts $N$ photons. This coupling enables to correlate the cavity with the qubit, by driving it through one of the two microwave ports $a$ and $b$, and further to extract work in an autonomous manner (Fig. 1).

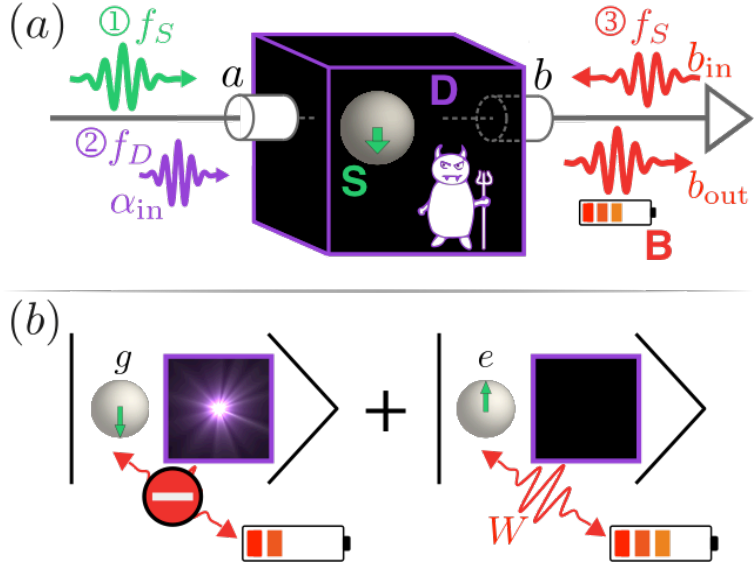

**Fig. 1. Sketch of the quantum Maxwell demon experiment.** (a) After preparation ① in a thermal or quantum state by a pulse at frequency $f_S$, the system S (superconducting qubit) state is recorded ② into the demon's quantum memory D (microwave cavity). In practice, a pulse incoming towards port $a$ at $f_D$ populates the cavity mode with a state $\rho_{\alpha_{in}}$ only if the qubit is in the ground state $|g\rangle_S$. This information is used to extract work $W$ ③, which charges a battery B (a microwave pulse at frequency $f_S$ on port $b$) with one extra photon. Importantly the system emits this photon only when the demon's cavity is empty. The work $W$ is determined by amplifying and measuring the average output power at $f_S$ on $b_{out}$. The memory reset step ④ is performed by cavity relaxation. (b) When the system starts in a quantum superposition of $|g\rangle_S$ and $|e\rangle_S$, the demon and system are entangled after step ②.

We now discuss the steps of the work extraction cycle. During step ① we prepare the system in a thermal state at an arbitrary temperature $T_S \geq T_S^0$ (Fig. 1), where $T_S^0 = 103 \pm 9$ mK is the equilibrium system temperature in the dilution refrigerator. This is realized by driving, in a fraction $p(T_S)$ of all experimental sequences, the qubit with a resonant $\pi$-pulse, which flips the qubit state, thus simulating thermalisation with a heat bath (see supporting materials [21]). Conveniently, this technique can also prepare non-thermal quantum states, such as an equal superposition of the qubit, by driving it with a $\pi/2$-pulse.

Step ② consists in encoding the state of the system into the demon's memory, which starts in the vacuum $|0\rangle_D$. Driving port $a$ with a pulse of amplitude $\alpha_{in}$ at frequency $f_D$ (Fig. 1) excites the demon's memory conditioned on the system being in $|g\rangle_S$. This requires the pulse duration to be longer than $\chi^{-1}$ and shorter than the coherence times of the qubit and cavity [21]. By design, decoherence of both system and demon's memory is dominated by spontaneous emission into port $b$ with respective relaxation rates $\gamma_S = (2.2 \text{ μs})^{-1}$ and $\gamma_D = (207 \text{ ns})^{-1}$. If the system starts in an arbitrary superposition $c_g|g\rangle_S + c_e|e\rangle_S$, it becomes entangled with the demon (Fig. 1); ideally $c_g|g\rangle_S \otimes |\alpha\rangle_D + c_e|e\rangle_S \otimes |0\rangle_D$, where $|\alpha\rangle_D$ is a coherent state. In practice the qubit-induced nonlinearity and decoherence of the cavity lead to an impure memory state $\rho_{\alpha_{in}}$ instead of $|\alpha\rangle_D$. The average photon number $\bar{n} = \text{Tr}(d^\dagger d \rho_{\alpha_{in}})$ is determined by fitting the numerical result of the full master equation to match the experimentally obtained system state [21].

The work extraction occurs during step ③. A coherent $\pi$-pulse, playing the role of the battery $B$, is sent through port $b$ at frequency $f_S$ (Fig. 1). Without the demon, the qubit would deterministically absorb (emit) a quantum of energy $hf_S$ from (into) the battery, if it is initially in $|g\rangle_S$ ($|e\rangle_S$). Crucially the demon prevents this transfer when its memory has $N \geq 1$ photons, because then the pulse is off resonance by $-N\chi$. When the correlation between the demon's memory having no photons and the system being in $|e\rangle_S$ is perfect only stimulated emission is allowed and work is extracted from system to battery. However, when the correlation is not perfect, in particular when $\bar{n} \ll 1$, the demon sometimes erroneously lets the qubit absorb a quantum of energy from the battery.

The demon thus ends in a state with an entropy $S_D$ of at least the decrease of system entropy, and has to be reset to close the thermodynamic cycle [2]. In the final step ④ of this experiment we let the demon's memory thermalize with a second bath that has a low temperature ($72 \pm 13$ mK). So this demon can extract work in a cyclic manner but it does so using a second bath, thus behaving as a regular heat engine.

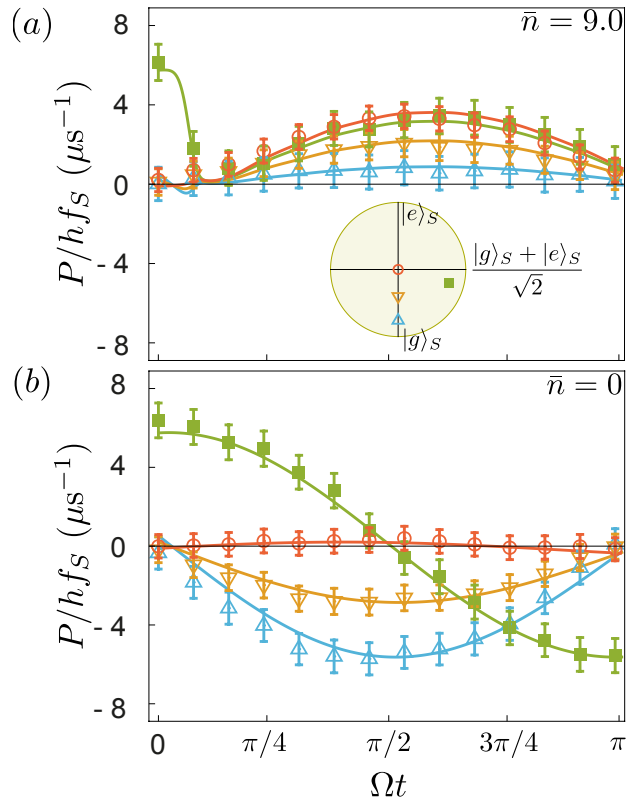

**Fig. 2. Measured extracted power.** Measured extracted power (normalized by a quantum of energy $hf_S$) for step ③ as a function of time during the pulse at $f_S$. Blue, orange and red symbols and error bars correspond to an initial thermalized system at temperatures $T_S = 0.17$, $0.40$ K and above 8K (see inset for initial Bloch vectors). Green symbols correspond to an initial quantum superposition obtained by a $3\pi/2$-pulse acting on the system at 0.10 K. Solid lines result from a numerical simulation with no fit parameters and match the measurements well. **(a)** The demon memory state $\rho_{\alpha_{in}}$ contains $\bar{n} = 9$ photons when encoding a system in $|g\rangle_S$. **(b)** Same figure for an ignorant demon ($\bar{n} = 0$ in step ②).

Remarkably, the power $P$ extracted from the system during step ③ when it is driven at $f_S$ can be directly accessed [21] through the difference between the incoming and outgoing photon rates of port $b$

$$\frac{P}{hf_S} = \langle b_{out}^\dagger b_{out}\rangle_B - \langle b_{in}^\dagger b_{in}\rangle_B = \gamma_b \frac{1+\langle\sigma_Z\rangle_S}{2} + \frac{\Omega}{2}\langle\sigma_X\rangle_S,$$

where $\sigma_X, \sigma_Y$ and $\sigma_Z$ are the Pauli matrices for the system. $\gamma_b$ is the Purcell coupling rate of the system to the transmission line through port $b$ and $\Omega \propto |\langle b_{in}\rangle_B|$ is the frequency of the Rabi oscillations around $\sigma_Y$ induced by the drive. The two contributions on the far right side can be identified as spontaneous emission of the system through port $b$ and stimulated emission. The latter is a coherent exchange of energy between the drive and system and as such contributes to the work extracted from the system.

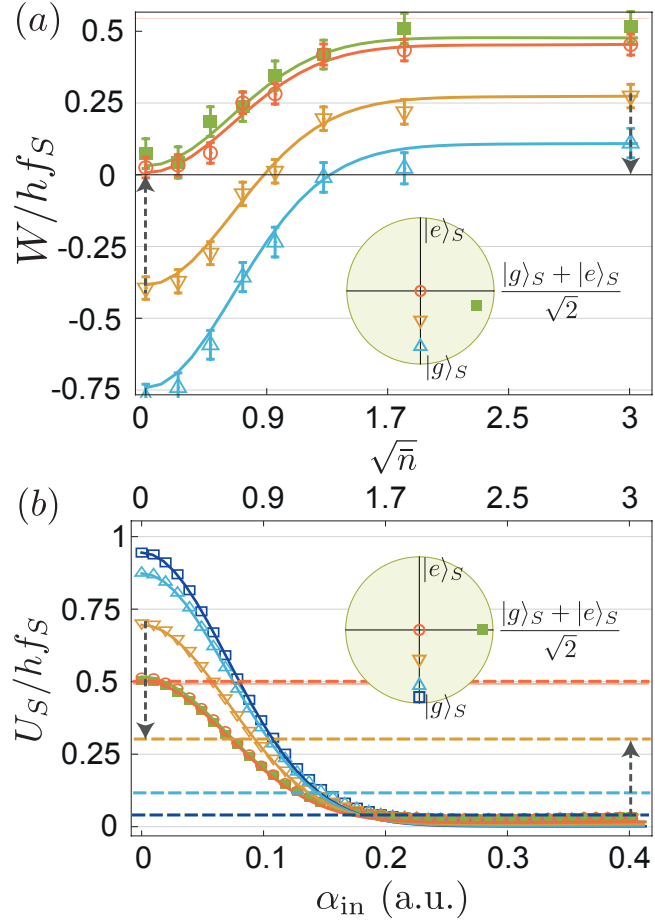

**Fig. 3. Work and internal energy of the system. (a)** Total extracted work during step ③ as a function of $\sqrt{\bar{n}}$, with $\bar{n}$ the number of photons in the demon memory. Symbols correspond to measurement of the battery and solid lines result from simulations. Colors correspond to the same initial states of the system as in Fig. 2. **(b)** Symbols: measured internal energy $U_S$ of the system at step ④ as a function of the amplitude $\alpha_{in}$ of the pulse at $f_D$ used in step ② to encode information in the demon's memory. Error bars are smaller than symbol size. Dashed lines indicate $U_S$ after preparation step ①. Solid lines result from the full master equation and establish the conversion between $\alpha_{in}$ and $\sqrt{\bar{n}}$. An additional dark blue color corresponds to an initial temperature $T_S = T_S^0$. The equality of extracted work and change in $U_S$ is highlighted by dashed arrows of identical lengths.

In the experiment, at $t = 0$ we send a pulse $b_{in}$ with a duration $\pi/\Omega$ with $\Omega = (67 \text{ ns})^{-1} \gg \gamma_S \geq \gamma_b$ so that spontaneous emission can be neglected and $\langle b_{out}^\dagger b_{out}\rangle_B - \langle b_{in}^\dagger b_{in}\rangle_B$ fully quantifies the work extraction. We measure the field intensity on $b_{out}$ using a near quantum limited heterodyne detection setup [21] to access directly the average instantaneous power extracted from the system [22]. The power is shown in Fig. 2 as a function of time during the pulse in step ③ in units of photons per microsecond for various initial system states (see inset) and for two values of the average photon number in the demon memory $\bar{n}$. In Fig. 2a the average photon number $\bar{n} = 9$ is large enough for the demon to distinguish the system states well. As expected from the demon's action, the measured power is positive for all initial states and greater for higher initial system temperature. In contrast when the demon is unable to distinguish $|g\rangle_S$ and $|e\rangle_S$, which happens for $\alpha_{in} = \bar{n} = 0$, the extracted power is measured to be negative for the system starting in any thermal state (Fig. 2b). This arises because the demon is ignorant and lets the system drain energy from the battery. This failure uncovers the role of information in the work extraction by the demon.

At $\bar{n} = 0$, a distinctive feature appears when the system starts in a quantum superposition of $|g\rangle_S$ and $|e\rangle_S$ (green). Even though the total work is zero, just like for the equally mixed state (red), the instantaneous power now oscillates illustrating the work potential of coherences. In contrast, for an efficient demon ($\bar{n} = 9$ in Fig. 2a), there is no quantum signature in the extracted work. Note that the peak in the green curve arises due to overlapping of steps ② and ③ to avoid transients [21].

Integrating the extracted power over the duration of step ③ gives the work $W = \int_0^{\pi/\Omega} P \, dt$, whose magnitude is at most $hf_S$ (Fig. 3a). As $\alpha_{in}$ increases, the demon's encoding improves and the work increases from negative to positive values. This extracted work is given by the change in the system internal energy $U_S = hf_S \langle e|\rho_S|e\rangle$ during step ③, $W = -\Delta U_S - Q \approx -\Delta U_S$, where $Q$ is the heat arising from spontaneous emission, which is negligible. While the work was measured on the battery, we independently measure $U_S$ as a function of $\alpha_{in}$ (Fig. 3b) at the end of step ④ using the cavity as a dispersive detector [20,21]. The variations of work (Fig 3a) indeed mirror the change of internal system energy between steps ① (dashed lines) and ④ (symbols). As $\alpha_{in}$ increases, the demon extracts more energy from the system making it end up close to the ground state (residual excitation of $2.7 \pm 1\%$) whatever the initial state (Fig. 3b). Indeed the thermodynamic cycle can be used to cool down superconducting qubits in practice, as previously demonstrated in its continuous version [23]. The full decay of $U_S$ and the increase of $W$ as a function of $\alpha_{in}$ are well reproduced numerically (solid lines in Fig. 3). It is natural to compare the extracted work with Landauer's work cost of erasure, $k_B T \ln 2$ [2,4]. Since the system is connected to a thermal bath only during step ①, the work extraction is not optimal. Indeed $W$ is limited by the initial internal energy $U_S = hf_S/(1 + \exp(hf_S/k_B T))$, so that it reaches at most 40% of the Landauer cost.

A key signature of Maxwell's demon is the transfer of entropy from the system to the memory [2]. In contrast to previous realizations of Maxwell demons [8–11], our experiment not only allows a direct measurement of work but also gives full access to the density matrix $\rho_D$ of the demon's memory, including its von Neumann entropy $S_D = -\text{Tr}(\rho_D \ln \rho_D)$. We perform a full quantum tomography of $\rho_D$ using the qubit as a measurement apparatus right after step ③ [21,24]. When the qubit starts close to $|g\rangle_S$, the Maximum Likelihood reconstruction [21,25] of the demon's state gives $\rho_{\alpha_{in}}$, which is entropic and far from a coherent state when $\alpha_{in} = 0.25$ as expected (Fig. 4a). In contrast, when the system starts in $|e\rangle_S$, the memory stays close to $|0\rangle_D$ with a small residual entropy (Fig. 4b).

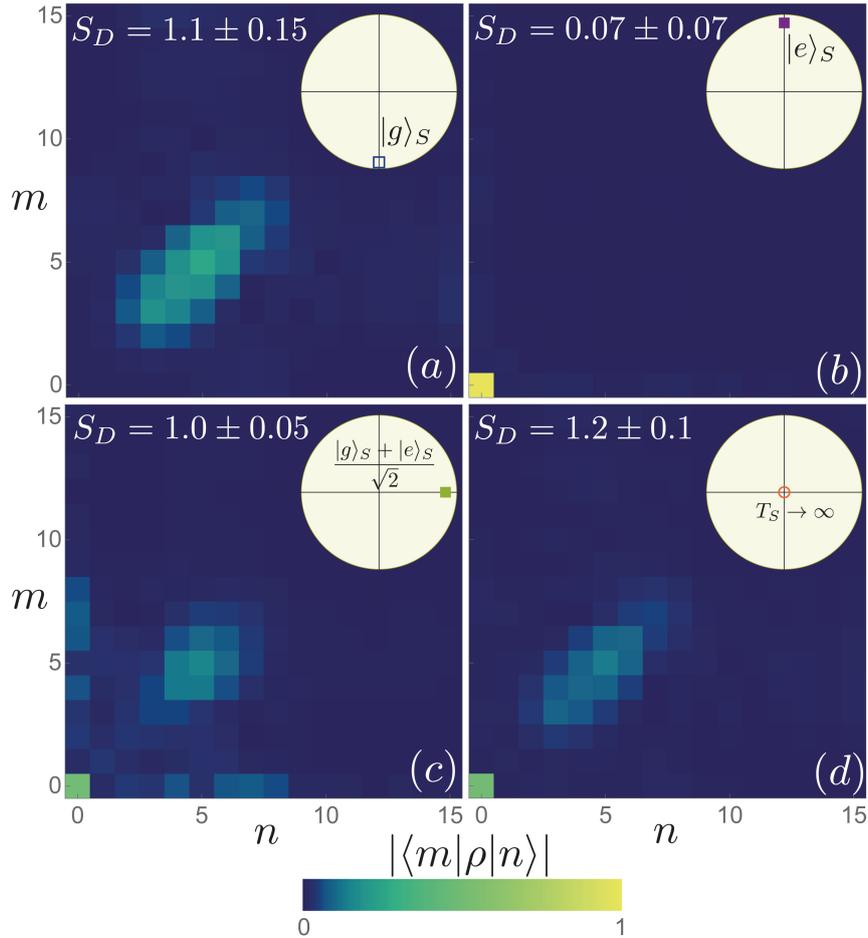

**Fig. 4. Tomography of the demon state.** Reconstructed density matrix $\rho_D$ by Maximum Likelihood at the end of the work extraction step ③ for $\alpha_{in} = 0.25$ and when the system is initially **(a)** at temperature 0.10 K, **(b)** close to the excited state, **(c)** a superposition of ground and excited states, **(d)** a maximally mixed state (see Bloch vector in insets). Each pixel represents the amplitude of a density matrix element in the Fock basis and the von Neumann entropies $S_D$ are given. (Wigner function shown in [21]).

Most interesting is the comparison of the effect on the demon when the system starts in a quantum superposition (Fig. 4c) or in a thermal state at large temperature (Fig. 4d). In the first case $S_D = 1.0 \pm 0.05$ and $\rho_D$ exhibits coherences between $|0\rangle$ and higher Fock states, while coherences are missing in the second case leading to a larger entropy $S_D = 1.2 \pm 0.1$. This transfer of non-classicality from the system to the memory is a signature of the quantum Maxwell demon. While the entropies of these two states are ordered as expected, their values are much larger than a bit of entropy, $\ln 2 \approx 0.7$. This is quantitatively reproduced by simulations [21] and arises because dissipation and non-linearity of the memory results in encoding in a large number of energy levels rather than in just two dimensions. Using a full tomography of the system [21], we have checked that the memory entropy $S_D$ is always higher than the system entropy decrease $S_S(①) - S_S(③)$.

Future developments of this experiment could involve superconducting circuits with a widely tunable frequency, which would allow the implementation of optimal quasistatic processes, where the system stays in equilibrium. This would allow a test of Landauer's principle in the quantum regime. The encoding fidelity of the demon is quantified by the mutual information between system and demon. By adding an extra qubit and readout cavity, one could demonstrate the expected proportionality between extracted work and mutual information [26–28]. Finally, with the level of control shown in the experiment, superconducting circuits provide an exciting platform to explore single shot quantum thermodynamics [29] and quantum heat engines [30].

**Acknowledgments**

We thank M. Devoret, P. Degiovanni, E. Flurin, Z. Leghtas, F. Mallet, V. Manucharyan, J. Pekola, J.M. Raimond, M. Ueda and the late M. Clusel for fruitful discussions and feedback. Nanofabrication has been made within the consortium Salle Blanche Paris Centre. This work was supported by the ANR under the grants 12-JCJC-TIQS and 13-JCJC-INCAL, by Ville de Paris through the grant Qumotel of the Emergence program and by the COST network MP1209 "Thermodynamics in the quantum regime". J.A. acknowledges support from EPSRC, grant EP/M009165/1, and the Royal Society.

# Supplementary Information for the letter
## Observing a quantum Maxwell demon at work

N. Cottet[1*], S. Jezouin[1*], L. Bretheau[1]†, P. Campagne-Ibarcq[1]‡, Q. Ficheux[1], J. Anders[2], A. Auffèves[3], R. Azouit[4,5], P. Rouchon[4,5] and B. Huard[1,6]

[1]Laboratoire Pierre Aigrain, Ecole normale supérieure, PSL Research University, CNRS, Université Pierre et Marie Curie, Sorbonne Universités, Université Paris Diderot, Sorbonne Paris-Cité, 24 rue Lhomond, 75231 Paris Cedex 05, France.
[2]University of Exeter, Stocker Road, Exeter EX4 4QL, United Kingdom.
[3]Institut Néel, UPR2940 CNRS and Université Grenoble Alpes, avenue des Martyrs, 38042 Grenoble, France
[4]Centre Automatique et Systèmes, Mines ParisTech, PSL Research University, 60 Boulevard Saint-Michel, 75272 Paris Cedex 6, France.
[5]Quantic Team, INRIA Paris-Rocquencourt, Domaine de Volucéau, B.P. 105, 78153 Le Chesnay Cedex, France.
[6]Laboratoire de Physique, Ecole Normale Supérieure de Lyon, 46 allée d'Italie, 69364 Lyon Cedex 7, France.

*these authors contributed equally to the work presented
†current address: Department of Physics, Massachusetts Institute of Technology, Cambridge, Massachusetts 02139, USA.
‡current address: Department of Applied Physics and Physics, Yale University, New Haven, Connecticut 06520, USA.

### 1/ Experimental setup

The qubit studied in this experiment follows the superconducting 3D transmon design. It consists of a single aluminum Josephson junction connected to two antennas forming a large capacitance shunting the junction, so that it behaves as a weakly anharmonic resonator. The antennas couple to the electromagnetic field, so that the system can be driven with microwaves. By addressing only the transition between the ground and first excited states, we isolate an effective qubit at $f_S = 7.088$ GHz with anharmonicity $\alpha = 126$ MHz.

The transmon is embedded in an aluminium cavity, anchored at the base-temperature (20 mK) of a dilution refrigerator (Fig. **S0**). The system is probed via two transmission lines terminated with SMA connectors whose central pin dips inside the cavity. The cavity and qubit energy damping rate are dominated by the coupling to the "out" transmission line. The first cavity mode (TE110, resonating at $f_D = 7.913$ GHz) dominates the electromagnetic environment of the transmon. Indeed, the transmon chip is placed at the maximum of the TE110 field and its antennas are aligned with the field polarization, whereas the next cavity mode has an amplitude node at the transmon position.

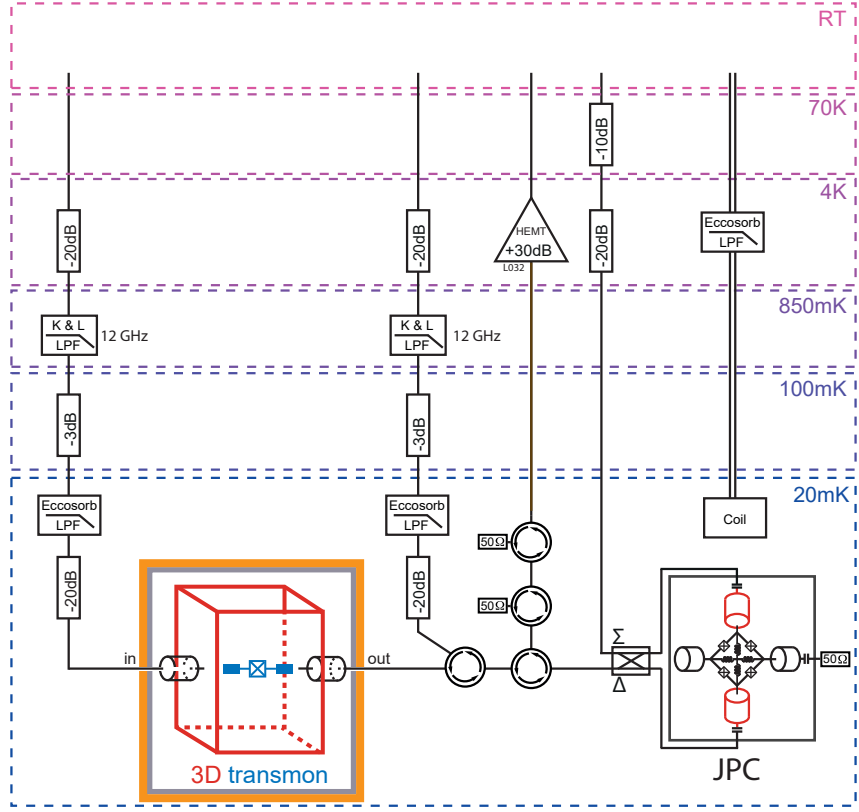

**Fig. S0.** Schematic of the wiring inside the dilution refrigerator.

2/ System entropy

Here we provide supplementary data investigating the link in between the demon's action and the resulting system entropy. Fig. **S1** shows the system entropy after step ④ as a function of $\alpha_{in}$. The former is determined by performing full tomography of the system density matrix, $\rho_S$, and using $S_S = -\text{Tr}(\rho_S \ln \rho_S)$. The system entropy directly reflects the predictability of the demon's action on the system and exhibits a pure quantum feature. Indeed, $S_S$ presents a maximum as a function of $\alpha_{in}$ when the demon's action maximally decreases the purity of the system state. At this maximum, the system and demon are maximally entangled after step ④ and measuring the state of the system only while discarding the state of the demon destroys the system purity. When one is only interested in the effect of the demon on the system this effect must be interpreted as an unpredictability of the demon's action, due to its quantum nature. This unpredictability can be characterized by noticing that the demon always sets the system in $|g\rangle$ when it starts in $|e\rangle$, but randomly excites the system in $|e\rangle$ with a probability $1 - \langle 0|\rho_{\alpha_{in}}|0\rangle$ when it starts in $|g\rangle$. Thus, at low temperature, the random behavior of the demon directly reflects on $S_S$ (blue), which evolves as $H(\langle 0|\rho_{\alpha_{in}}|0\rangle)$, where $H(p) = -p \log_2 p - (1-p)\log_2(1-p)$ denotes the Shannon entropy. In contrast, the entropy of the system at large temperature (red) can only decrease as the demon puts it in a state going monotonously from $(|g\rangle\langle g| + |e\rangle\langle e|)/2$ to $|g\rangle$ as $\alpha_{in}$ increases and $\langle 0|\rho_{\alpha_{in}}|0\rangle$ decreases. From Fig. 3b of the main text, it is clear that although there is no difference between the final internal energy of a system initially at large temperature (red) or in an equal quantum superposition of $|g\rangle$ and $|e\rangle$ (green), the entropy of these classical and quantum cases strongly differ. In particular, the entropy of a system starting in a quantum superposition never reaches 1 bit in contrast with any of the classical cases. Again, the simulation of the full master equation (solid lines) matches well the measured entropy in all cases.

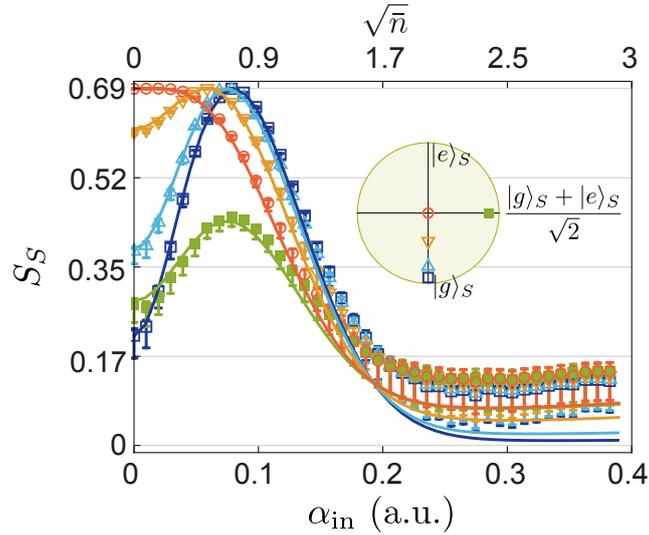

**Fig. S1. System entropy.** Symbols and error bars: measured system entropy $S_S$ after step ④ as a function of $\alpha_{in}$ for various initial states (inset, see also the main text). Solid lines: simulations.

### 3/ Modeling and characterizing the system-demon evolution

This part presents the theoretical model that we use to describe the combined system-demon and how the physical parameters entering the model are measured.

#### 3.1) System-demon evolution

The system-demon Hamiltonian reads:

$$\frac{H}{h} = f_S|e\rangle\langle e| + f_D d^\dagger d - \chi d^\dagger d |e\rangle\langle e| - K(d^\dagger d)^2 + \chi_2 (d^\dagger d)^2 |e\rangle\langle e| + H_{\text{drive}}(t)/h \ .$$

The first three terms correspond to the standard dispersive Hamiltonian of cQED and are similar to the Hamiltonian described in the main text, with $f_S = 7.088$ GHz, $f_D = 7.913$ GHz and $\chi = 33.8$ MHz (see 1.2 for the parameters measurement). The two following terms are small – though important – corrections which are more than one order of magnitude smaller than $\chi$ in magnitude. The first one describes the demon self non-linearity due to the Josephson junction (self-Kerr) with $K = 0.7$ MHz. This term is responsible for a photon-number dependent resonance frequency of the demon. A direct consequence is that a coherent drive on the demon does not build a coherent state as it would with a linear oscillator. The second one represents the first non-linear correction of the dispersive shift $\chi$ and reads $\chi_2 = 0.9$ MHz. The frequency of the time-dependent driving Hamiltonian is adapted accordingly to the system and demon effective frequencies.

The time evolution of the global system and demon density matrix $\rho$ is given by the master equation

$$\frac{d\rho}{dt} = -\frac{i}{\hbar}[H(t),\rho] + \kappa_D \mathcal{L}(d)\rho + \gamma_\downarrow \mathcal{L}(\sigma_-)\rho + \gamma_\uparrow \mathcal{L}(\sigma_+)\rho + \frac{\gamma_\phi}{2}\mathcal{L}(\sigma_z)\rho \qquad (S1)$$

with $\hbar = h/2\pi$, $\mathcal{L}(\hat{O})\rho$ the Lindblad superoperator $\mathcal{L}(\hat{O})\rho = \hat{O}\rho\hat{O}^\dagger - \frac{1}{2}(\rho\hat{O}^\dagger\hat{O} + \hat{O}^\dagger\hat{O}\rho)$ ; $\sigma_- = |g\rangle\langle e|$ the lowering operator from $|e\rangle$ to $|g\rangle$ ; $\sigma_+ = |e\rangle\langle g| = \sigma_-^\dagger$ ; and $\sigma_z = |e\rangle\langle e| - |g\rangle\langle g|$ the Pauli matrix along $z$. The first Lindblad term represents the energy decay of the demon with a rate $\kappa_D/2\pi = 0.77$ MHz. The second (resp. third) term represents the system relaxation (resp. excitation) with rate $\gamma_\downarrow = (1 - p_e^0)\gamma_1$ (resp. $\gamma_\uparrow = p_e^0\gamma_1$), where $\gamma_1 = (2.2~\mu s)^{-1} = 454$ kHz and $p_e^0 = 3.6\%$. The last term represents pure dephasing of the system with rate $\gamma_\phi = 85$ kHz.

### *3.2) Parameter estimation*

The bare qubit frequency $f_S = 7.088$ GHz is measured to high accuracy by standard Ramsey experiment. The photon-number-dependent qubit frequency is then measured with two-tone spectroscopy, as represented Fig. **S2**. An initial displacement of the cavity leads to several peaks at frequencies $f_S^n$ in the qubit spectrum corresponding to different Fock states of the cavity according to $f_S^n = f_S - n(\chi + n\chi_2)$. Fitting the peaks center frequencies up to 6 photons, we obtain the dispersive shift $\chi = 33.8$ MHz and its non-linearity $\chi_2 = 0.9$ MHz. The two small peaks in the spectroscopy at $f_S \pm 65$ MHz correspond to small leakage of the modulation pulse on the cavity line.

Measuring the ac-Stark shift and measurement-induced dephasing of the system when the cavity is weakly driven around its resonance frequency gives access to the cavity frequency and loss rate. We drive the cavity at frequency $f_D + \Delta$ and amplitude $\epsilon_d$ weak enough to build a coherent state $|\alpha_g(t)\rangle$ (resp. $|\alpha_e(t)\rangle$) when the qubit is in $|g\rangle$ (resp. $|e\rangle$). In the frame rotating at $2\pi f_D$ they obey the following equations:

$$\dot{\alpha}_g(t) = 2i\pi\Delta\alpha_g(t) - \frac{\kappa_D}{2}\alpha_g(t) + \epsilon_d$$
$$\dot{\alpha}_e(t) = 2i\pi(\Delta + \chi - \chi_2)\alpha_e(t) - \frac{\kappa_D}{2}\alpha_e(t) + \epsilon_d .$$

Due to the photons stored in the cavity the qubit frequency is detuned by an amount $f_{\text{Stark}}(t)$ and the qubit acquires an extra dephasing rate $\Gamma_d(t)$ given by (*1*)

$$f_{\text{Stark}}(t) = (\chi - \chi_2)\text{Re}[\alpha_g^*(t)\alpha_e(t)]$$
$$\Gamma_d(t) = (\chi - \chi_2)\text{Im}[\alpha_g^*(t)\alpha_e(t)] .$$

The above expressions are valid in the case of a qubit coupled to a linear cavity. In our experiment the cavity non-linearity cannot be neglected at high drive amplitude. To avoid this problem we drive the cavity such that $\alpha_{g,e} \approx 0.5$ at resonance ($\Delta = 0$ and $\Delta = -\chi + \chi_2$). The ac-Stark shift $f_{\text{Stark}}$ and measurement-induced dephasing $\Gamma_d$ are determined by Ramsey fringes experiments. Fig. **S3** shows their dependence in $\Delta$. Fitting the data with the theoretical expressions gives $f_D = 7.913$ GHz, $\kappa_D = (T_D)^{-1} = 0.77$ MHz and $\chi - \chi_2 = 33.1$ MHz (in agreement with the qubit spectroscopy). The good agreement between the experimental points and the theoretical curves ensures that the corrections due to the cavity Kerr term could be neglected in this case.

The value of the Kerr coefficient is determined by coherently exciting the cavity and then measuring the probability to find it in vacuum $\langle 0|\rho_{\alpha_{\text{in}}}|0\rangle$. We drive the cavity with a 100 ns-long drive of variable amplitude $\alpha_{\text{in}}$ followed by a $\pi$-pulse of the qubit conditioned on the cavity being in the vacuum (similar to ②). The subsequent probability to find the qubit in the excited state $P(|e\rangle)$ is therefore a measurement of $\langle 0|\rho_{\alpha_{\text{in}}}|0\rangle$. We find (Fig. **S4**) coherent oscillations that are a generalization of Rabi oscillations to a weakly non-linear oscillator (6). They are very well reproduced by solving numerically the full master equation (*S1*) with $K = 0.7$ MHz. This large Kerr term compared to usual transmon-cavity experiments is due to the relatively small qubit-cavity detuning. In particular, we have $K \approx \kappa_D$, possibly leading to strong deformations of the cavity spectroscopy from a Lorentzian shape.

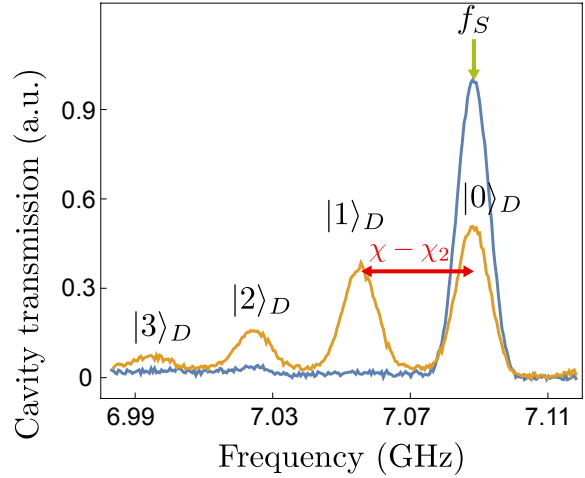

**Fig. S2. Qubit two-tone spectroscopy** with cavity (blue) at thermal equilibrium and (orange) after displacement. Each peak corresponds to one cavity Fock state. The frequency coordinate indicates the frequency of a $\pi$-pulse performed just before measuring the cavity transmission.

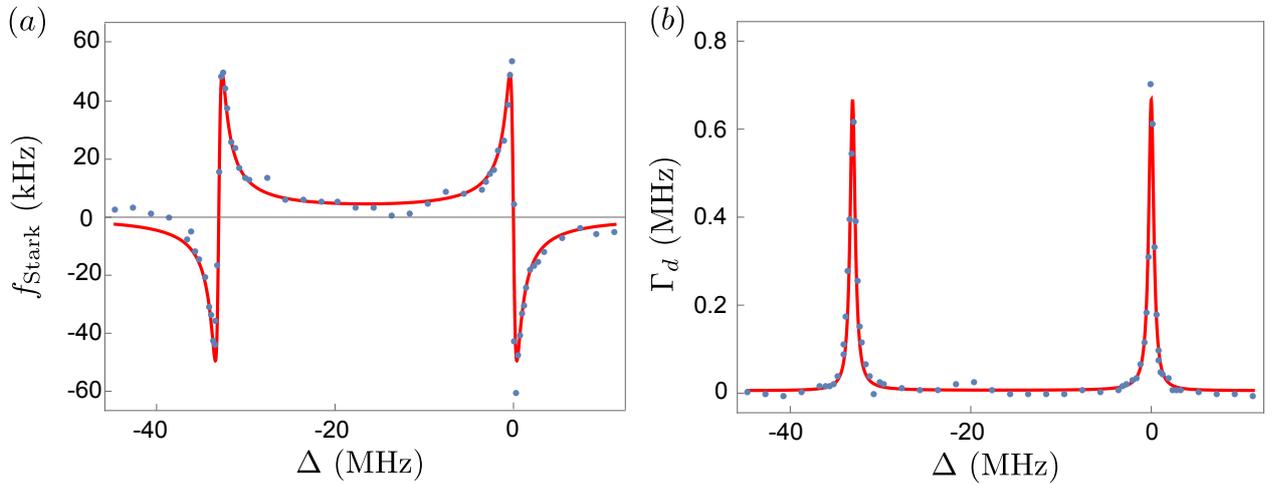

**Fig. S3. (a) AC-Stark shift and (b) measurement-induced dephasing** of the qubit when the cavity is weakly driven at $f_D + \Delta$. Points are experimental data and solid lines are fits.

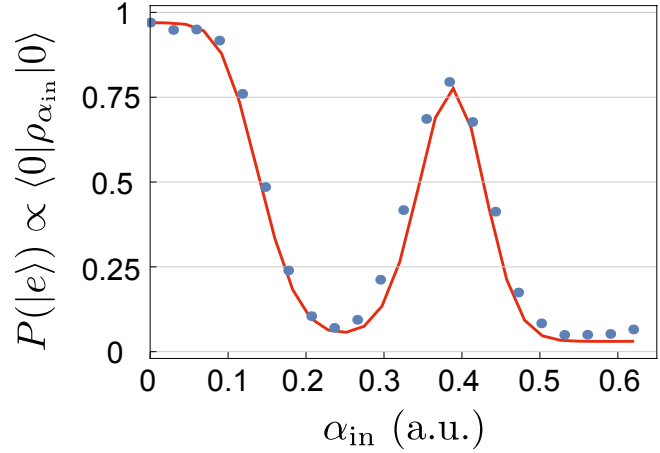

**Fig. S4. Rabi-like oscillations of the cavity and Kerr estimation.** Points show the measured probability to find the qubit in $|e\rangle$ after a 100ns-long drive of the cavity with amplitude $\alpha_{in}$ followed by a $\pi$-pulse of the qubit conditioned on the cavity being in the vacuum. Solid line is obtained by solving numerically the full master equation with $K = 0.7$ MHz.

### *3.3) Temperature measurement*

The demon temperature is estimated with the qubit spectroscopy at thermal equilibrium (blue curve Fig. S2). The relative heights of peaks corresponding to different Fock states $|n\rangle_D$ are given by the relative probabilities for the demon to contain $n$ photons. Writing $\rho_D^0$ the density matrix of the demon when it is at thermal equilibrium with its bath at temperature $T_D^0$, those probabilities are given by the Maxwell-Boltzmann distribution

$$P(n) = \langle n|\rho_D^0|n\rangle = \frac{e^{-n\frac{hf_D}{k_B T_D^0}}}{Z}$$

with $Z = \sum_{n=0}^{\infty} e^{-nhf_D/k_B T_D^0}$ the partition function. Comparing the peaks at $n = 0$ and $n = 1$ gives $P(1) = 0.7 \pm 0.5\%$ at thermal equilibrium, which corresponds to a temperature $T_D^0 = 72 \pm 13$ mK.

The large Kerr term makes difficult to measure the system's temperature by standard cavity spectroscopy means with a sufficiently good accuracy. Instead we measured the contrast difference between two sets of Rabi oscillations of the system with different initial states. To do so we used the second excited state $|f\rangle_S$. The resonance frequency between states $|e\rangle_S$ and $|f\rangle_S$ is equal to $f_S - E_C/h = 6.962$ GHz where $E_C$ is the charging energy of the transmon. We drive the system at $f_S - E_C/h$ to obtain Rabi oscillations between $|e\rangle_S$ and $|f\rangle_S$ and measure the transmission of the cavity at $f_D - 2(\chi - \chi_2)$, the resonance frequency of the demon when the system is in $|f\rangle_S$. The first set of oscillations is obtained by driving the system at thermal equilibrium and the second one is obtained by applying a $\pi$-pulse between $|g\rangle_S$ and $|e\rangle_S$ before driving the system. The two sets of Rabi oscillations are represented in Fig. S5. At thermal equilibrium with its bath at temperature $T_S^0$ the system is represented by the density matrix $\rho_S^0$ and the contrast of the oscillations is given by $C^{eq} = \langle e|\rho_S^0|e\rangle - \langle f|\rho_S^0|f\rangle$. When $k_B T \ll hf_S$ the population in $|f\rangle$ is negligible and we simply get

$$C^{eq} \approx \langle e|\rho_S^0|e\rangle = \frac{1}{1+e^{\frac{hf_S}{k_B T_S^0}}}.$$

The initialization $\pi$-pulse on the second set exchanges the populations in $|g\rangle_S$ and $|e\rangle_S$ with a fidelity $F_\pi$ and the contrast of the oscillations is then given by

$$C^\pi = F_\pi \left(1 - \frac{1}{1 + e^{\frac{hf_S}{k_B T_S^0}}}\right) + \frac{1 - F_\pi}{1 + e^{\frac{hf_S}{k_B T_S^0}}}$$

$$C^\pi = \frac{1 + F_\pi \left(e^{\frac{hf_S}{k_B T_S^0}} - 1\right)}{1 + e^{\frac{hf_S}{k_B T_S^0}}}.$$

The ratio $C^\pi/C^{eq}$ hence gives a direct measurement of the quantity $F_\pi e^{hf_S/k_B T_S^0}$. The fidelity is estimated to be $F_\pi = 92\%$ from the known system decoherence rate and pulse durations. This yields to

$$T_S^0 = 103 \pm 9 \text{ mK}$$

which corresponds to an equilibrium population in $|e\rangle_S$ of $p_e^0 = 3.6 \pm 1\%$. This 1% uncertainty on the system residual excitation takes into account the temperature fluctuations measured during the time of the experiment.

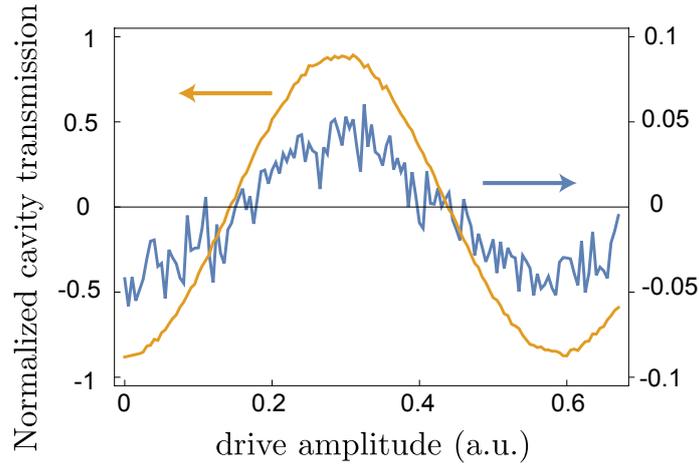

**Fig. S5. System temperature measurement by Rabi oscillations.** Two Rabi oscillations are measured using the cavity transmission at $f_D - 2(\chi - \chi_2)$, (blue) at thermal equilibrium and (orange) after a $\pi$-pulse between the states $|e\rangle_S$ and $|g\rangle_S$. The blue curve is magnified 10 times. The y-axis represents the real part of the transmission of the cavity, which is normalized so that 1 indicates a state in $|f\rangle$, while $-1$ indicates that there is zero probability to find the system in $|f\rangle$.

### *3.4) System-demon simulation*

All the theoretical curves presented in the main text (Fig. 2 and 3) are obtained by numerically solving equation (*S1*) for the whole pulse sequences in a Hilbert space truncated to 45 photons. To do so we have used the Python library "Qutip" (*4*). Due to signal filtering and absorption between the microwave sources and the cavity the effective amplitudes of the drives are only known up to a scaling factor. Numerically we solve this problem by calibrating the pulses at $f_D$ and $f_S$. More precisely, simulating Rabi oscillations of the qubit by varying the pulse amplitude at constant length calibrates the $\pi$-pulse amplitude. On the other hand the full simulation of the demon sequence yields the system internal energy at state ④ as a function of the numerical displacement amplitude. The simulated evolution is then rescaled to match the experimental data by a simple proportionality factor (see Fig. 3b of the main text). This calibration gives directly the nonlinear relationship between $\alpha_{\text{in}}$ and $\sqrt{n}$.

### 4/ Pulse sequences

In this part we describe in detail the pulse sequences realizing the Maxwell's demon experiment, as well as the tomography of the system and demon.

We actually used two distinct pulse sequences to realize the Maxwell's demon experiment. On the one hand, the sequence should be performed as fast as possible in order to minimize the effect of decoherence and relaxation of the qubit and the cavity. This improves the demon efficiency and allows to keep trace of quantum signatures of the experiment such as negativities in the demon Wigner function. On the other hand, the tiny work performed by the demon is detected by a JPC quantum limited amplifier which suffers limited bandwidth, and thus sets a minimum duration for the work extraction.

The first sequence (Fig. **S6a**), called "sequential" in the following, is a fast sequence that consists in the three steps described in the main text. This sequence has been used for the results presented in Fig. 3b and 4 of the main text. The system preparation (step ①) is made using, if needed, a Gaussian-shaped $\pi$- or $\pi/2$- pulse at $f_S$ with a standard deviation of 12.5 ns and truncated at ±25 ns. Step ② is done by displacing the demon by a Gaussian-shaped pulse at $f_D$ with a standard deviation of 12.5 ns and truncated at ± 25 ns. Work extraction (step ③) is finally performed using a $\pi$-pulse with the same parameters as in step ①.

The second demon sequence (Fig. **S6b**), called "continuous", allows to measure the work performed by the demon and is the one used for the results presented in Figs. 2 and 3a of the main text. In this version a single long continuous square pulse at $f_S$ realizes steps ① and ③. Its Rabi frequency is $\Omega/2\pi = (416 \text{ ns})^{-1} \ll \text{BW}_{\text{JPC}}$ (see below). The demon displacement (step ②) is performed on top of this long pulse and defines the passage from step ① to ③. It is Gaussian-shaped with a shortened standard deviation of 10 ns (truncated at ± 20 ns) which is much smaller than the Rabi period. The starting time of the displacement determines the initial state of the qubit. We use 200, 300 and 400 ns starting times, which realize approximately $\pi$-, $3\pi/2$- and $2\pi$- pulses during step ①, respectively. As will appear below, the reason why we do not stop the pulse at $f_S$ during step ② and perform a $2\pi$- (resp. $3\pi/2$-) pulse instead of nothing (resp. a $3\pi/2$-pulse) to prepare the low temperature state (resp. an equal superposition of ground and excited states) is to minimize the impact of the transient response of the JPC on the work measurement. All this result in a slightly smaller purity of the system when the "continuous" sequence is used, as compared to the "sequential" one (compare the insets of Fig 3a and b of the main text).

Those demon sequences are followed by various measurement sequences depending on the quantity that we want to measure. The system tomography (Fig. **S6c**) is realized by first letting the demon relax towards thermal equilibrium during $3 \mu s$, then performing (eventually) a $\pi/2$ rotation around $x$ or $y$, and finally measuring the population in $|e\rangle$ with standard High Power Readout measurement (*3*). The demon tomography (Fig. **S6d**) will be described in part 6.

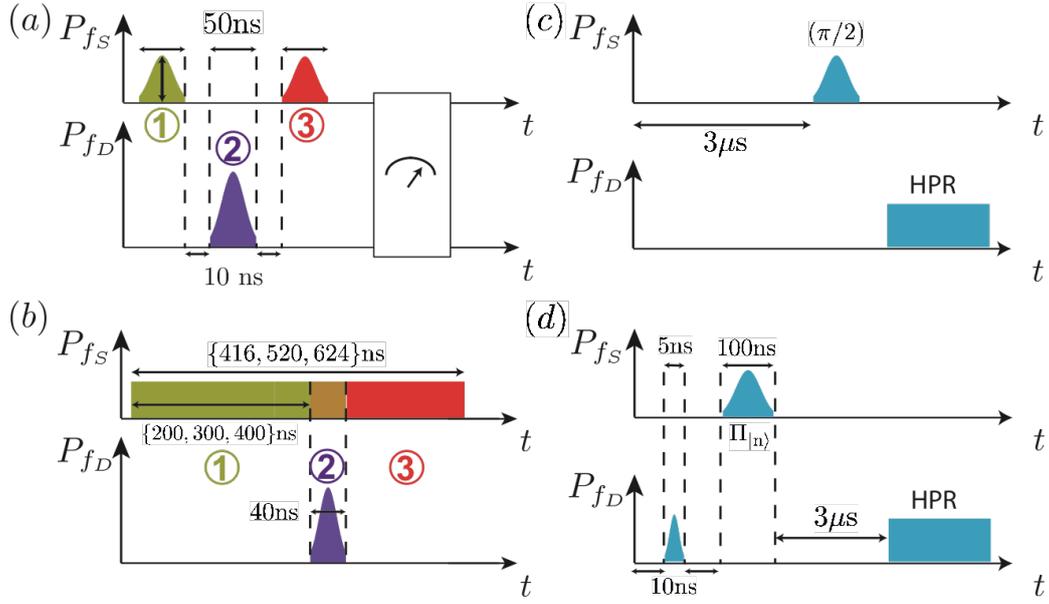

**Fig. S6. Pulse sequences.** (a) "Sequential" and (b) "continuous" sequences realizing the Maxwell's demon experiment. (c) Qubit and (d) cavity tomography pulse sequences.

5/ Photon rate measurement

The purpose of this part is to describe in detail how the work performed by the qubit by stimulated emission is measured and the different calibrations necessary to make this measurement.

### *5.1) Photons radiated by the system*

The electromagnetic field propagating in port b at the system frequency is decomposed in two parts : an input (resp. output) field $b_\text{in}$ (resp. $b_\text{out}$) that propagates towards (resp. away from) the qubit and cavity. The so-called input-output formalism gives the following relation (*5*)

$$b_\text{out} = b_\text{in} - \sqrt{\gamma_p}\sigma_-$$

where $\gamma_p$ is the Purcell rate of the coupling of the qubit with port b. Multiplying this equation by its hermitian conjugate leads to the output photon rate

$$b_\text{out}^\dagger b_\text{out} = b_\text{in}^\dagger b_\text{in} + \gamma_p \sigma_+\sigma_- - \sqrt{\gamma_p}(b_\text{in}^\dagger \sigma_- + b_\text{in}\sigma_+).$$

The expectation value of the photon rate is obtained by tracing over both the qubit and transmission line degrees of freedom:

$$\langle b_\text{out}^\dagger b_\text{out}\rangle = \langle b_\text{in}^\dagger b_\text{in}\rangle_b + \gamma_p \frac{1+\langle\sigma_Z\rangle_S}{2} - \sqrt{\gamma_p}\langle b_\text{in}^\dagger \sigma_- + b_\text{in}\sigma_+\rangle_{S,b}$$

where we have used the identity $\sigma_+\sigma_- = |e\rangle\langle g||g\rangle\langle e| = |e\rangle\langle e| = (1+\sigma_Z)/2$. The drive being classical we can assimilate the operator $b_\text{in}$ to a simple complex number $\beta_\text{in}$. Since the drive phase defines the phase of the measurement, we can further assume that $\beta_\text{in}$ is a real negative number. One can then show that the qubit oscillates around the *y*-axis of the Bloch sphere with the Rabi frequency $\Omega/2\pi = \sqrt{\gamma_b}|\beta_\text{in}|/\pi$ The photon rate thus reads

$$\langle b_{out}^\dagger b_{out}\rangle = |\beta_{in}|^2 + \gamma_p \frac{1+\langle\sigma_Z\rangle_S}{2} - \sqrt{\gamma_p}\beta_{in}\langle\sigma_- + \sigma_+\rangle_S$$
$$\langle b_{out}^\dagger b_{out}\rangle = |\beta_{in}|^2 + \gamma_p \frac{1+\langle\sigma_Z\rangle_S}{2} + \frac{\Omega}{2}\langle\sigma_X\rangle_S$$

with $\sigma_X = |e\rangle\langle g| + |g\rangle\langle e| = \sigma_- + \sigma_+$ the Pauli matrix along *x*. The first term simply corresponds to the photons of the incident drive that are reflected off the cavity. The second term is equal to the Purcell rate times the probability to find the qubit in the excited state ; this is the spontaneous emission. In the regime $\Omega \gg \gamma_1, \gamma_\phi$ it is fully incoherent and can thus be interpreted as *heat* emitted by the qubit. To understand the last term, consider a $\pi$-pulse applied to the qubit. In the absence of decoherence, one can easily show that:

$$\int_{|e\rangle}^{|g\rangle} \frac{\Omega}{2}\langle\sigma_X\rangle_S \mathrm{d}t = 1$$

The qubit has thus emitted one photon on top of the input drive. Hence this term represents stimulated emission ; this is the *work* provided by the qubit to the drive.

### *5.2) Heterodyne measurement*

The fluorescence field emitted through port *b* is first amplified by a homemade Josephson Parametric Converter (JPC) followed by a commercial amplifiers. Then it is down-converted to 62.5 MHz, digitized and numerically demodulated to get the in-phase and out-of-phase signals $(I(t), Q(t))$ with a sampling timestep $dt = (62.5\text{ MHz})^{-1} = 16$ ns. We have the following relationship in between the demodulated signals and the output field and power:

$$\overline{I(t)} = \sqrt{G}\times\mathrm{Re}(\langle b_{\mathrm{out}}(t)\rangle)$$
$$\overline{Q(t)} = \sqrt{G}\times\mathrm{Im}(\langle b_{\mathrm{out}}(t)\rangle)$$
$$\overline{I(t)^2 + Q(t)^2} = \text{offset} + G\times\langle b_{out}^\dagger b_{out}\rangle \ .$$

The overlines indicate averaging over $10^7$ measurement records ; the offset term in the last equation is due to the technical noise of the *I* and *Q* measurements ; *G* is the overall (power) gain of the amplification chain. Using a vector network analyzer we determine the gain and bandwidth of the JPC: $G_{\mathrm{JPC}} \approx 14$ dB and $\mathrm{BW}_{\mathrm{JPC}} = 29$ MHz. This working point is chosen such that the bandwidth is much larger than the Rabi frequency of the qubit during the demon sequence. In the frequency domain this means that the JPC gain is almost constant on a range of frequency corresponding to the Mollow triplet, and in the time domain this means that the JPC correlation time is much smaller than the Rabi period.

### *5.3) Photon number calibration*

To access the output photon rate $\langle b_{out}^\dagger b_{out}\rangle$ (thus the work performed by the system) one has to calibrate the overall amplification chain gain, $G$. To do so we continuously drive the qubit at resonance through port *b* for various drive amplitudes resulting in various Rabi frequencies. The system performs damped Rabi oscillations with

$$\langle\sigma_Z\rangle_S(t) = \langle\sigma_Z\rangle_S^0 \cos(\Omega t)e^{-t/T_R}$$
$$\langle\sigma_X\rangle_S(t) = \langle\sigma_Z\rangle_S^0 \sin(\Omega t)e^{-t/T_R}$$
$$\langle\sigma_Y\rangle_S(t) = 0$$

where $T_R$ is the damping time of Rabi oscillations and $\langle\sigma_Z\rangle_S^0 = \text{Tr}(\sigma_Z\rho_S^0)$ is the initial thermal state of the qubit along $z$. The measurement outcomes therefore read

$$\overline{I(t)} = \sqrt{G}\beta_{in} + \sqrt{G}\times\sqrt{\gamma_p}\langle\sigma_Z\rangle_S^0\sin(\Omega t)e^{-t/T_R}$$
$$\overline{Q(t)} = 0$$
$$\overline{I(t)^2 + Q(t)^2} = \text{offset} + G|\beta_{in}|^2 + G\times\langle\sigma_Z\rangle_S^0\frac{\gamma_p[1+\cos(\Omega t)] + \Omega\sin(\Omega t)}{2}e^{-t/T_R}.$$

$\overline{I}$ and $\overline{I^2+Q^2}$ exhibit damped oscillations with $\Omega$-dependent amplitudes given by

$$A^I(\Omega) = \sqrt{G}\times\sqrt{\gamma_p}|\langle\sigma_Z\rangle_S^0|$$
$$A^{I^2+Q^2}(\Omega) = G\times|\langle\sigma_Z\rangle_S^0|\sqrt{\gamma_p^2 + \Omega^2}\Big/2 \xrightarrow[\Omega\gg\gamma_p]{} G\times|\langle\sigma_Z\rangle_S^0|\Omega/2$$

Since we already know the qubit temperature, we can extract the gain $G$ by measuring the slope $dA^{I^2+Q^2}/d\Omega$. Fig **S7a** shows $\overline{I^2+Q^2}$ for three different Rabi frequencies. The oscillations can be strongly deformed at small times due to the transient response of the JPC to the start-up of $\beta_{in}$. The oscillations are thus fitted at high enough times. Fig **S7c** and **d** show in blue points the fitted values of $A^I$ and $A^{I^2+Q^2}$ as a function of the Rabi frequency. While a constant $A^I$ is expected, a linear reduction is observed with increasing $\Omega$. This reduction is considerably smaller when the qubit is driven in transmission through the *a* port instead of the *b* port (green points). The reduction can have two origins : saturation of the JPC due to the large $|\beta_{in}|$ and filtering of the signal due to the finite bandwidth $\text{BW}_{\text{JPC}}$. Since the transmission measurement is only sensitive to filtering effects, we deduce that filtering is negligible, as expected. We therefore model the reduction of $A^I(\Omega)$ when the qubit is driven trough the *b* port by using a $|\beta_{in}|$-dependent (or equivalently $\Omega$-dependent) amplitude gain (black line) $\sqrt{G(\Omega)} = \sqrt{G_0}(1-(\Omega/\Omega_\infty))$, where $G_0$ is the gain at zero input power and $\Omega_\infty$ is some constant. This $G(\Omega)$ also explains very well the saturation of $A^{I^2+Q^2}$ at high $\Omega$ (black line, Fig **S7d**) and therefore directly gives the correspondence between the measured power $\overline{I^2+Q^2}$ and the output photon rate $\langle b_{out}^\dagger b_{out}\rangle$.

Finally, we mention that the gain of the amplification setup underwent some small drift between the photon rate calibration and the realization of the experiment. It is represented by the red dots on Fig. **S7c** and **d** that give the amplitude and Rabi frequency of oscillations measured during the demon experiment. This drift was taken into account in the figures of the main text.

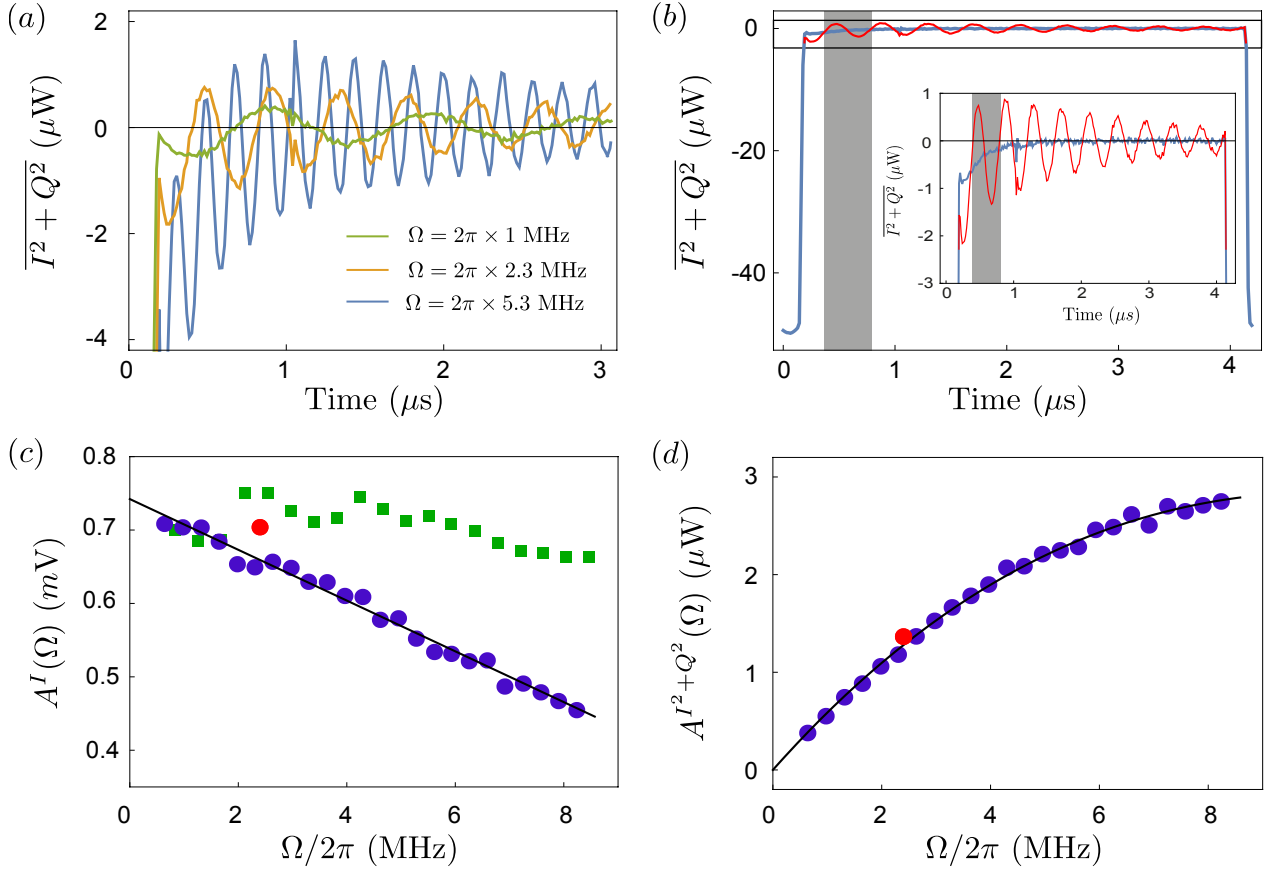

**Fig. S7. Photon number calibration.** (a) Photon rate measurement of the qubit under a continuous Rabi drive for three driving amplitudes. (b) Transient response of the JPC to the start-up of $\beta_{in}$. (c) $\Omega$-dependence of the amplitude of the Rabi oscillations measured on $I$ (blue points). The black curve is a fit with a linear gain model (see text). The red dots correspond to the Rabi oscillations used in the demon experiment and the green squares to Rabi oscillations measured in transmission. (d) Same for $I^2 + Q^2$.

### 5.4) Correction of the transient response of the JPC

As already noticed in Fig **S7a** the transient response of the JPC to the start-up of $\beta_{in}$ leads to significant deformation of $\overline{I^2 + Q^2}$ at small time. To correct for this deformation during the work measurement in the "continuous" sequence (step ③ in Fig. **S6b**), we do the following. First, we measure $\overline{I^2 + Q^2}$ (red curve, Fig **S7b**) during a long drive of the qubit whose amplitude is such that the Rabi frequency precisely equals that of the "continuous" sequence, that is $\Omega/2\pi = (416\ \text{ns})^{-1}$. Second, we fit these oscillations at high enough time so that the JPC has reached its stationary regime. Finally, we extrapolate the fitted function down to Time=0μs and remove it from the measured $\overline{I^2 + Q^2}$. The obtained curve is shown in blue Fig. **S7b**. It represents the JPC transient response and is removed from the raw data in order to obtain Fig. 2 of the main text. For information, the grey rectangle Fig. **S7b** represent the moments where step ③ happens.

### 6/ Tomography of the demon state

In this part, we present the raw measurements and the procedure which allowed to obtain the density matrices of the demon shown Fig. 4 of the main text.

### *6.1) Demon Husimi Q-function*

The quantum state $\rho_D$ of the demon after step ③ can be fully characterized by the set of the generalized Husimi Q-functions :

$$Q_n(\beta) = \frac{1}{\pi} \langle n | D(\beta)^\dagger \rho_D D(\beta) | n \rangle.$$

In the absence of relaxation, they would be measured by the sequence represented Fig. **S6d**. The pulse at $f_c$ is so short that it performs a unitary displacement $\beta$ of the cavity independently of the self-Kerr $K$. The subsequent $\pi$-pulse is centered at $f_q - n\chi + n^2\chi_2$ and is chosen long enough so that it is conditioned on the cavity being in $|n\rangle$. The final measurement of the qubit excitation probability $p_{n,\beta}^e$ thus gives the probability to find $n$ photons in the displaced cavity, hence $Q_n(\beta)$. The relaxation however disturbs this measurement and has to be fully taken into account in order to reconstruct $\rho_D$ from the final measurements $p_{n,\beta}^e$.

We measured $p_{n,\beta}^e$ for $0 \leq n \leq 5$ and $-5.95 \leq \beta \leq 5.95$ discretized on a 31×31 square array. Raw measurements are shown Fig. **S8** (top line) when $\alpha_{in} = 0.25$ and the initial state of the system is (a) at temperature 0.10 K, (b) close to the excited state and (c) a superposition of ground and excited states. Note that the qubit and the cavity are disentangled at the beginning of the tomography. The measurements agree well with calculations (bottom line) obtained by solving numerically the full master equation *(S1)*. The numerical calculations include the full time-dependent pulse sequences of Fig. **S6a** and **d**.

### *6.2) Maximum Likelihood reconstruction*

We now describe how $\rho_D$ is reconstructed from the set of final measurements $p_{n,\beta}^e$.

First, we need to find the coefficients $k_e, k_0, k_\beta$ relating the voltages $V_{\text{Alazar}}$ (measured by the acquisition board) and $V_{\text{awg}}$ (generating the displacement pulse) to the in situ quantities $p_{n,\beta}^e = k_e V_{\text{Alazar}} + k_0$ and $\beta = k_\beta V_{\text{awg}}$. To do so, we perform (Fig. **S9**) a tomography of the thermal state $\rho^0 = \rho_D^0 \otimes \rho_S^0$ and simply match the experimental data (points) with the calculations (lines). The later are given by $p_{n,\beta}^e = \text{Tr}(\rho^0 E_{n,\beta}(0))$, where the operators $E_{n,\beta}$ describe the time-reversed system-demon evolution during the tomography pulse sequence represented Fig. **S6d** :

$$E_{n,\beta}(T) = I_D \otimes |e\rangle\langle e|$$

$$\frac{dE_{n,\beta}(t)}{dt} = -\frac{i}{\hbar}[H_{n,\beta}(t), E_{n,\beta}(t)] - \kappa_D \bar{\mathcal{L}}(d) E_{n,\beta}(t) - \gamma_1 \bar{\mathcal{L}}(\sigma_-) E_{n,\beta}(t) - \frac{\gamma_\phi}{2} \bar{\mathcal{L}}(\sigma_z) E_{n,\beta}(t)$$

where $\bar{\mathcal{L}}(\hat{O})E = \hat{O}^\dagger E \hat{O} - \frac{1}{2}(E\hat{O}^\dagger\hat{O} + \hat{O}^\dagger\hat{O}E)$. The subscript of the Hamiltonian $H$ reminds that the qubit drive frequency and cavity drive amplitude depend on $n$ and $\beta$, respectively. There is a very good agreement in between data and theory. The main effect of relaxation here is to widen and shift the peaks to higher $\beta$.

To reconstruct the state $\rho_D$ after step ③, we use a MaxLike method (7). Assuming Gaussian noise of the measurement records, $\rho_D$ is expected to maximize the log-likelihood function :

$$f(\rho_D) \propto -\sum_{n,\beta} \left(p_{n,\beta}^e - \text{Tr}(\rho_D \otimes \rho_S E_{n,\beta})\right)^2,$$

where $\rho_S = p_g|g\rangle\langle g| + (1-p_g)|e\rangle\langle e|$ is the state of the qubit after the Maxwell demon sequence. From Fig. **S1**, we estimate that $p_g \approx 0.97$. Maximization of $f(\rho_D)$ is achieved by a gradient algorithm with orthogonal projection on the (convex) subspace of matrices that are positive, hermitian and with unit trace.

All the physical parameters entering the MaxLike algorithm are measured independently. Two technical parameters are however also needed. First the infinite Hilbert space of the cavity is truncated so that only Fock states such as $n \leq N_{\text{trunc}}$ are taken into account. Second we do not use all the displacements for the reconstruction but only those lower than a bound $|\beta| \leq \beta_{max}$. Clearly, one should have $N_{\text{trunc}} \gg \beta_{max}^2 > \bar{n}$. But on the other hand, $N_{\text{trunc}}$ and $\beta_{max}$ should not be too big since the dispersive Hamiltonian in (S1) is only valid for low photon number. Fig. **S10** shows the dependence of the demon's entropy on $N_{\text{trunc}}$, $\beta_{max}$ and $p_g$ for a qubit initially in (a) a temperature 0.1K, (b) a superposition of ground and excited states, (c) close to the excited state and (d) a maximally mixed state. Fig. **S10a** and **b** show that the truncation should be taken on the plateau $13 \leq N_{\text{trunc}} \leq 21$, while there is a negligible dependence on $p_g$. The uncertainty on the entropy comes mostly from the choice of $\beta_{max}$ (Fig. **S10d**). We find $S_D(b) = 1.0 \pm 0.05$ and $S_D(d) = 1.2 \pm 0.1$. In comparison, the simulations shown Fig. **S8** predict $S_D(b) = 1.06$ and $S_D(d) = 1.17$.

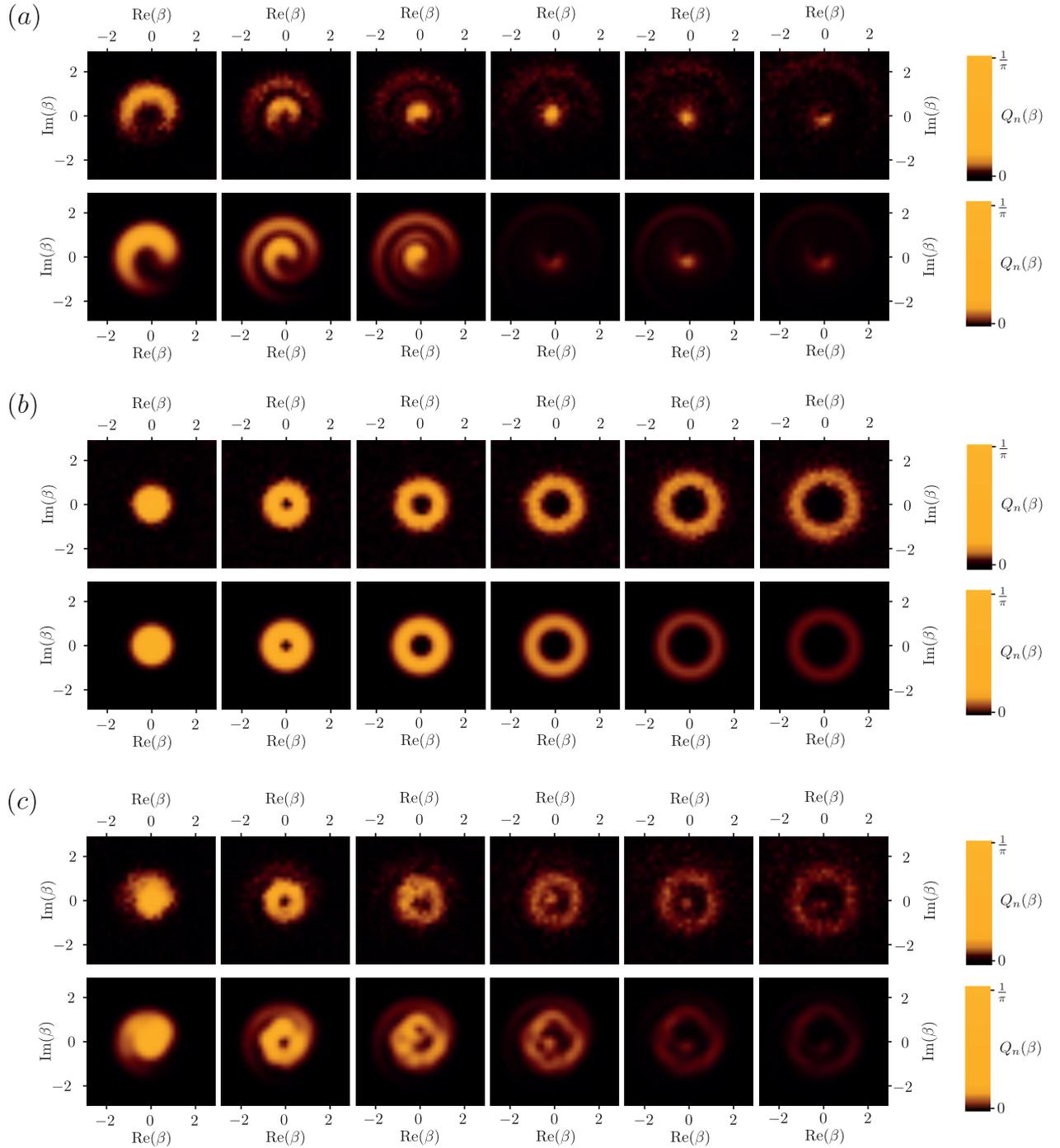

**Fig. S8. Generalized Husimi Q-functions of the demon.** Experimental (top line) and calculated (bottom line) Husimi Q-functions of the demon's memory after step ③ for $\alpha_{in} = 0.25$ and the system initially close to (a) $|g\rangle_S$, (b) $|e\rangle_S$ and (c) a quantum superposition of $|e\rangle_S$ and $|g\rangle_S$. From left to right, n=0,1,2,3,4,5.

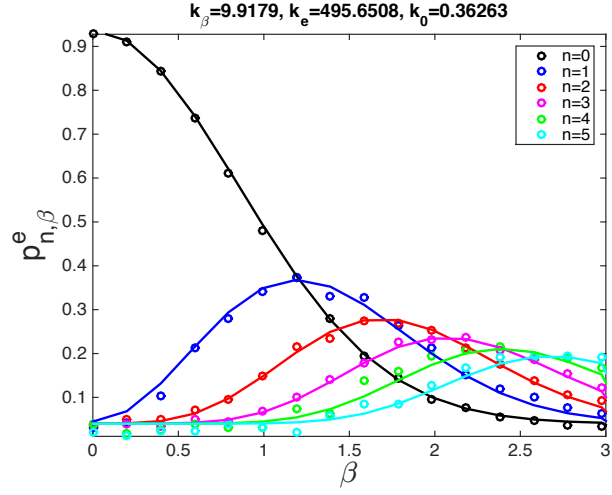

**Fig. S9. Tomography of the thermal state $\rho^0$.** Points are measurements rescaled with the values of $k_e, k_0, k_\beta$ indicated on top of the figure. Lines are calculations performed on an Hilbert space with $N_{\text{trunc}} = 15$.

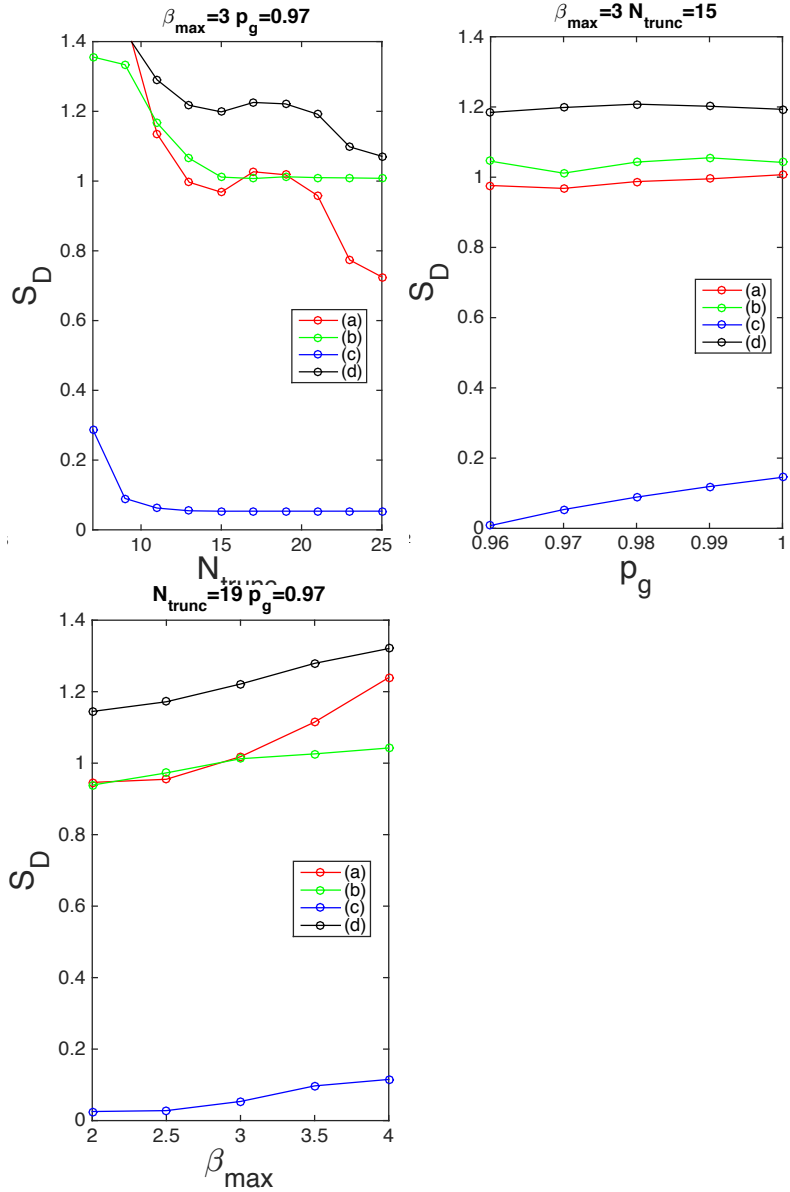

**Fig. S10. Uncertainty on the demon entropy due to the MaxLike reconstruction.** The demon entropy, $S_D = -\text{Tr}(\rho_D \ln \rho_D)$, calculated from the reconstructed demon density matrix, $\rho_D$, is plotted for various values of the parameters $N_{\text{trunc}}$, $\beta_{max}$ and $p_g$ entering the MaxLike algorithm. The qubit was initially prepared in (a) a temperature 0.1K, (b) a superposition of ground and excited states, (c) close to the excited state and (d) a maximally mixed state.

7/ Effective thermal bath generation

The temperature $T_S$ of a qubit in state $\rho$ can be defined by the Boltzmann weight $\langle e|\rho|e\rangle = \left(1 + e^{\frac{hf_S}{kT_S}}\right)^{-1}$.

$T_S$ is indeed a properly defined temperature in the sense that the qubit will be in a state $\rho$ such as above when coupled to an environment at temperature $T_S$. In fact, the temperature of a qubit only depends on the ratio between its emission and absorption rates, which is entirely set by the quantum noise at frequency $f_S$ since (8,9) $e^{-hf_S/kT_S} = S[-f_S]/S[f_S]$.

The relaxation rate of the qubit into the environment also depends on this spectral density as $T_1^{-1} \propto S[f_S] + S[-f_S]$.

Since only the quantum spectral density at $\pm f_S$ matters, they are many ways to create an effective heat bath for the qubit. The choice of thermal environment is guided by a tradeoff between proximity to an actual heat bath coupled to the qubit and controllability of the bath.

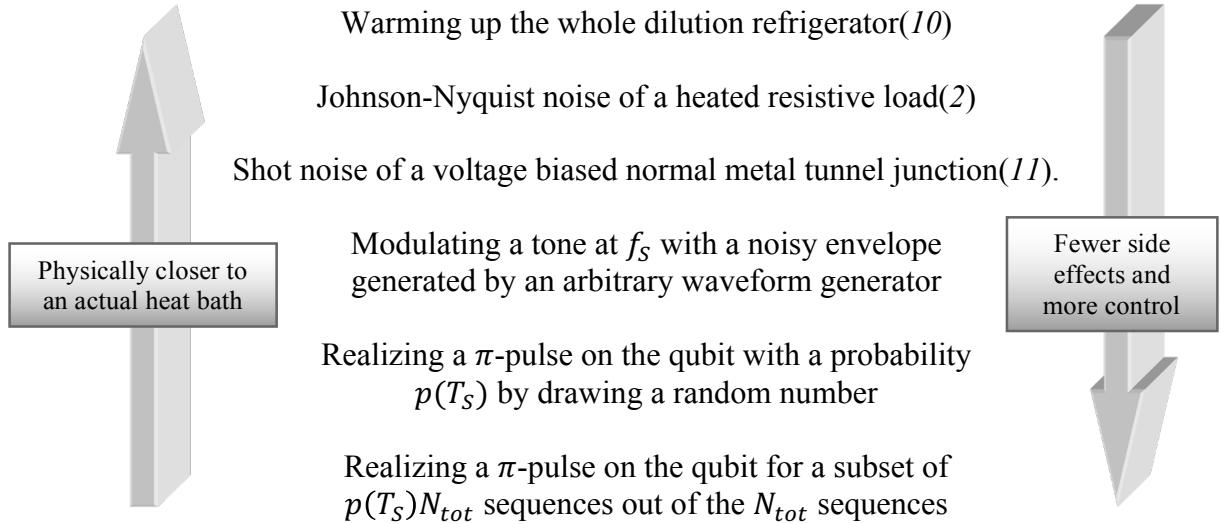

Physically closer to an actual heat bath

Warming up the whole dilution refrigerator(*10*)

Johnson-Nyquist noise of a heated resistive load(*2*)

Shot noise of a voltage biased normal metal tunnel junction(*11*).

Modulating a tone at $f_S$ with a noisy envelope generated by an arbitrary waveform generator

Realizing a $\pi$-pulse on the qubit with a probability $p(T_S)$ by drawing a random number

Realizing a $\pi$-pulse on the qubit for a subset of $p(T_S)N_{tot}$ sequences out of the $N_{tot}$ sequences

Fewer side effects and more control

Since the effect of all these heat baths is identical, we chose to use the most practical one, which consists in driving the qubit with a $\pi$-pulse during the preparation stage for only a fraction $p(T_S)$ of the experimental sequences and leave it in the initial equilibrium state for a fraction $1 - p(T_S)$. The probability $p(T_S)$ has to be chosen such that for an initial qubit temperature $T_S^0$ and for a fidelity $F_\pi$ of the $\pi$-pulse the correct Boltzmann weight at $(T_S)$ is reproduced,

$$\frac{1}{1+e^{hf_S/kT_S}} = p(T_S)\left(F_\pi\left(1 - \frac{1}{1+e^{\frac{hf_S}{kT_S}}}\right) + \frac{1-F_\pi}{1+e^{\frac{hf_S}{kT_S}}}\right) + \frac{(1-p(T_S))}{1+e^{\frac{hf_S}{kT_S}}}.$$

This determines the form of $p(T_S)$

$$p(T_S) = \frac{1}{F_\pi}\frac{\left(1+e^{\frac{hf_S}{kT_S^0}}\right)\left(1+e^{\frac{hf_S}{kT_S}}\right)^{-1} - 1}{e^{\frac{hf_S}{kT_S^0}} - 1}.$$